\documentclass[aps,showpacs,pre,twocolumn]{revtex4}

\usepackage{psfig}

\newcommand{\Sec}[1]{Section~\ref{#1}}     
     
\newcommand{\Fig}[1]{Fig.~\ref{#1}}

\begin{document}
\title{Static response in disk packings}

\author{Cristian F.~Moukarzel, Hector Pacheco-Mart\'{\i}nez,
  J.~C.~Ruiz-Suarez}

\affiliation{CINVESTAV del IPN, Depto.\ de F\'\i sica Aplicada, \\
  97310 M\'erida, Yucat\'an, M\'exico }

\author{Ana M. Vidales}

\affiliation{Depto.\ de F\'\i sica y CONICET, Universidad Nacional de San Luis
  \\ Chacabuco 917, 5700 San Luis, San Luis, Argentina }
\date{\today}
\begin{abstract}
  We present experimental and numerical results for displacement response
  functions in packings of rigid frictional disks under gravity. The central
  disk on the bottom layer is shifted upwards by a small amount, and the
  motions of disks above it define the displacement response. Disk motions are
  measured with the help of a still digital camera. The responses so measured
  provide information on the force-force response, that is, the excess force
  at the bottom produced by a small overload in the bulk. We find that, in
  experiments, the vertical-force response shows a Gaussian-like shape,
  broadening roughly as the square root of distance, as predicted by diffusive
  theories for stress propagation in granulates.  However, the diffusion
  coefficient obtained from a fit of the response width is ten times larger
  than predicted by such theories.  Moreover we notice that our data is
  compatible with a crossover to linear broadening at large scales.  In
  numerical simulations on similar systems (but without friction), on the
  other hand, a double-peaked response is found, indicating wave-like
  propagation of stresses. We discuss the main reasons for the different
  behaviors of experimental and model systems, and compare our findings with
  previous works.
\end{abstract}
\maketitle
\section{Introduction}
\label{sec:introduction}
Stress distributions in static granular materials display puzzling
characteristics~\cite{JNBG96,WGF97,GR98,GG99b} that do not quite fit into
classical elastic descriptions, and have defied attempts at analytic modeling
for some time already.  The observation of a pressure dip below conical piles,
force chains, sudden macroscopic changes in stress patterns under slight
perturbations, and exponential (instead of Gaussian) stress distributions,
among other phenomena, have triggered intense theoretical and experimental
work. As a result, a multiplicity of stress propagation models have been put
forward.  The $q$-model~\cite{LNSF95,CLMM96} assumes diffusive behavior for
the vertical stress component considered as a scalar quantity, and gives rise
to an exponential distribution of stresses.  Other scalar models in turn
predict Gaussian~\cite{SSSF99} or power-law distributed~\cite{CBS97} stresses.
By postulating a linear relation between stress components~\cite{BCCS95}, a
wavelike equation~\cite{WCCA96} is derived for stress propagation, the so
called OSL model~\cite{WCCS97}.  This model reproduces the pressure
dip~\cite{WCCA96}, and is consistent with stresses in silos~\cite{VCBS00}.  A
memory formalism~\cite{KSPN98,KS01} contains as special limits the wavelike
and diffusive behaviors. Furthermore, a recent description in terms of
scattering force-chains~\cite{BCLF01} gives rise to wavelike propagation on
small scales, crossing over to something similar to classical elasticity on
larger scales.
\\
Linear elasticity describes the propagation of stresses in terms of
differential equations of the \emph{elliptic} type, wavelike propagation
corresponds to the \emph{hyperbolic} case, while diffusive behavior is the
borderline, or \emph{parabolic} case. These three descriptions give rise to
very different responses~\cite{GR98,GG99b} when a small force is applied on a
localized region on the upper surface of a packing. Linear elasticity predicts
a bell-shaped response, having a width proportional to depth.  A diffusive
behavior, on the other hand, implies that the width of the response scales as
the square root of depth.  Finally, a wavelike propagation would be evidenced
by a response that is maximum on a diffuse annulus of linearly growing radius
(the ``light-cone'') in three dimensions, or by a response showing two
diverging peaks in two dimensions.
\\
One might expect such differences to be easily resolved by properly designed
experiments. However, presently available experimental results are not
conclusive.  Some small-scale results support the validity of the
$q$-model~\cite{D-SRS00}, while recent experiments using photoelastic
techniques~\cite{GRCG03x} appear to be in conflict with a diffusive picture.
The memory formalism has been shown to reproduce the stress oscillations
observed in laterally confined packings~\cite{SKHN98}.  Experiments on
sand~\cite{RCG01,SRCS01} show a single-peaked response function whose width
scales linearly with depth, as predicted by elasticity.  However the precise
shape of the response does not quite match~\cite{SRCS01} that of a linear
elastic medium. It has been noted that even systems in the elliptic regime can
have two peaks in their response functions~\cite{OBCA03}. It appears at
present difficult, based on available experimental results, to clearly
validate, or disprove, any of the stress propagation models that have been
proposed.
\\
Numerical measurements on disordered packings of frictionless disks, that
respect the property of isostaticity~\cite{MI98,RG00}, both
on-lattice~\cite{MI01,MR02} (with contact disorder) and
off-lattice~\cite{TWS99,TWS00,HTWR01}, show two clearly distinguishable peaks
in their response functions. However in experiments wavelike (or hyperbolic)
response functions have only been observed in ordered packings~\cite{GHLF01},
and it has been argued that disorder produces a crossover to an elliptic
description on large scales.
\\
Notice that the measurement of response functions in granular packings poses a
subtle experimental challenge. A distinction must be made between the response
to \emph{infinitesimal} perturbations $G^i$, derived under the assumption that
the contact network does not change, and $G^f$, the response under small but
\emph{finite} perturbations, i.e.\ allowing for rearrangements. On disordered
isostatic packings, $G^i$ is singular~\cite{MI98,MI01,MR02} and does not have
a well defined continuum limit. This is so because the propagation of stresses
is described by random multiplicative processes on these systems. In practice
this implies that $G^i$ takes positive as well as negative values, whose
modulus grows \emph{exponentially} with distance from the point where the
perturbation is applied. Thus on isostatic disordered packings, any finite
perturbation, no matter how small, necessarily induces contact rearrangements,
because a large and negative Green function corresponds to a contact that will
open upon perturbation. This anomalous sensitivity to perturbation is due to
isostaticity, and has been suggested as being responsible for the tendency of
stiff packings to reorganize upon perturbation~\cite{MI98}. Based on very
general physical grounds, one can expect rearrangements to fundamentally
modify response functions. Thus one should expect experimentally measured
response functions to strongly depend on the magnitude of the perturbation
whenever isostaticity is satisfied.  Possible effects of rearrangements are
discussed for example in Ref.~\cite{RG00} (See also recent discussions in
Refs.~\cite{RC02,HTWR02}).
\\
Moreover, isostaticity has only been rigorously proven~\cite{MI98,RG00} for
\emph{frictionless} packings. It is known that friction gives rise to
indeterminacies~\cite{RBRN96,HEA99}, and this is not compatible with
isostaticity. In real packings friction is important, and it is at present not
clear whether isostaticity applies in some restricted sense, or not at all.
Recent molecular dynamics results~\cite{SEGG02} on frictional deformable
spheres are not compatible with the packing being isostatic, although previous
similar studies~\cite{MJSP00} reached different conclusions.
\\
Interparticle forces in granular packings have been previously measured by
means of carbon-paper experiments~\cite{LNSF95,JNBG96,MJNF98,BMMF01,MJNS02},
photoelastic techniques~\cite{GHLF01,GRCG03x}, pressure
sensors~\cite{SRCS01,RCG01}, and high-precision balances~\cite{LMFF99}. None
of these methods is optimal for the determination of response functions, a
task that requires the measurement of small forces over small regions. Several
of the methods used up to now require large forces to be applied to the
packing, while others are not able to detect forces on single particles but
only averages over relatively large regions.  The use of an electronic balance
gives very precise results, but requires an extremely time consuming scanning
of the bottom of the packing.
\\
In this paper we report our first results from a novel experimental technique
allowing precise measurements of response functions in two-dimensional
packings. Our procedure consists in measuring the displacement-displacement
response function~\cite{MI98,MI01,MR02}, that is, the displacement produced on
a given disk by an upward displacement of one of the disks on the lowermost
layer. For isostatic systems, it has been shown~\cite{MI98,MI01,MR02} that
this quantity is exactly equivalent to the stress-stress response function.
Several recent numerical studies make use of this equivalence to measure
stress responses~\cite{MI98,MI01,MR02,TWS99,TWS00,HTWR01}, however this is the
first time that this equivalence between stress-stress and
displacement-displacement response functions is used in experiments.
%%%%%%%%%%%%%%%%%%%%%%%%%%%%%%%%%%%%%%%%%%%%%%%%%%%%%%%%%%%%%%%%%%%%%%%
\begin{figure}[h]
\centerline{\psfig{figure=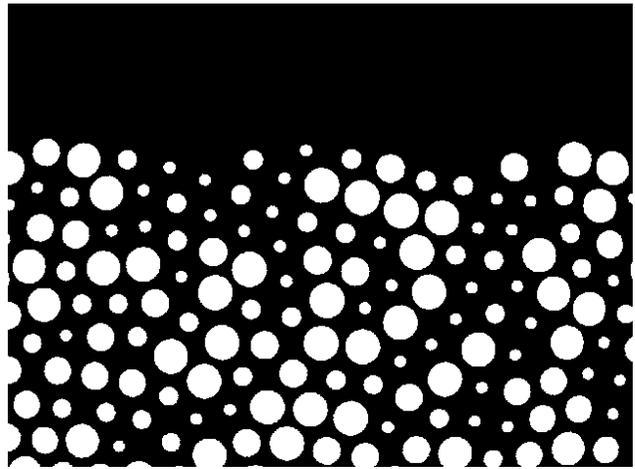,width=8.3cm,angle=0}}
\caption{{} A disk configuration as seen by the camera. The camera resolution
  is $640\times 480$ pixels. The approximate size of the viewable field is $29
  \times 22$ cm, implying a spatial resolution of roughly $0.45$ millimeter
  per pixel. Only around 150 disks out of a total of 400 in the container, fit
  in the camera view-field. The average number of layers is between nine and
  ten.}
\label{fig:disks}
\end{figure}
%%%%%%%%%%%%%%%%%%%%%%%%%%%%%%%%%%%%%%%%%%%%%%%%%%%%%%%%%%%%%%%%%%%%%%
\\
In standard response function measurements, a small force $f$ is applied on a
point on the top surface of a packing (defined as the origin of coordinates)
and the response function $G(x,y)$ is defined as the excess stress induced at
$\{x,y\}$, divided by $f$. In our experiments, the response function $G(x,y)$
gives the excess stress at $\{0,0\}$ (the central particle on the lowermost
row of the packings) produced by a small force $f$ acting at $\{x,y\}$. If
boundary effects can be neglected, these two definitions of $G(x,y)$ should
give statistically equivalent results. In other words, although for a given
sample these two ways to measure the response give different results, after
sample averaging one obtains the same function, if translation invariance
applies.
\\
Some advantages of the experimental procedure presented in this work are as
follows.  The measurement of displacements can be done with much better
precision and by simpler means than that of forces. Vertical and horizontal
displacements provide information respectively on vertical ($G_y$) and
horizontal ($G_x$) responses. We obtain information on $G_x(x,y)$ and
$G_y(x,y)$ for all $\{x,y\}$ in one measurement. Moreover, our technique does
not involve the subtraction of two force patterns, a procedure which is prone
to error, and more so when two large force patterns are subtracted to obtain a
small response. In our experiments, the displacement of each particle is a
direct measurement of the green function at that point.
\\
This work is organized in the following manner: \Sec{sec:experimental-device}
contains a description of the experimental setup used for the measurement of
response functions. In \Sec{sec:eresults} our results are presented and
analyzed. A comparison with numerical results obtained on frictionless
packings is established in \Sec{sec:numer-exper}. Finally, our results are
summarized and discussed in \Sec{sec:discussion}.
\section{Experimental setup}
\label{sec:experimental-setup}
\subsection{Experimental device}
\label{sec:experimental-device}
The experimental device consisted of a rectangular container made of two
parallel $615$mm by $320$mm plexiglass plates. The back and front plates were
respectively $34$ and $10$ mm thick, and they were separated by $5$ mm.  400
aluminum disks with a thickness of $4$ mm were confined between the plates.
These disks had diameters $16, 17, 18$ and $19$mm (100 of each).  The
container was placed vertically in order to minimize friction effects between
disks and walls.  All disks were painted black, and a white circular label was
affixed onto each of them in order to allow for motion measurement using
digital means. Labels had a slightly smaller radius than the disks they were
fixed on.
\\
The disks' positions and movements were recorded using a still digital camera
with a resolution of 640x480 pixels. The camera pointed to a rectangle of
$29\times 22$cm around the middle of the plexiglass container, implying a
resolution of roughly $0.45$mm per pixel. Before each experiment $i$, the
packing was shaken and allowed to settle under gravity alone. The disk
configuration before perturbation was then recorded using the camera,
obtaining a pre-image $I^0(i)$.  \Fig{fig:disks} shows a typical image. Next
the central disk in the lowest row was displaced upwards by 1mm. This disk was
fixed to a micrometric screw that fitted through a specially devised hole in
the bottom border of the container.  At this point the final configuration of
disks was recorded and stored to post-image $I^f(i)$. This procedure was
repeated a total of 330 times.
\\
A $C$ program processed these images to obtain individual disk motions. Our
program took pairs of images (before and after the perturbation) in b\&w
(1bpp) format as input. For each pre-image, all clusters of white pixels were
first identified by the method of burning.  Their geometric centers were taken
to be disks centers. These centers were then taken as starting points for the
identification (burning) of clusters on the corresponding post-image. This
allowed to find the displacement suffered by each disk in the pre-image. For
each disk, horizontal and vertical displacements $G_x$ and $G_y$ were
calculated from the differences between geometric centers. These numbers (two
per disk) give the displacement-displacement response function for a given
packing, which for frictionless systems equals the force-force response
functions, as discussed somewhere else~\cite{MI98,MI01,MR02}.  Because of the
existence of friction in our experiments, displacement responses are not
exactly equal to stress responses, however these differences will be regarded
as small and ignored in the following.
%%%%%%%%%%%%%%%%%%%%%%%%%%%%%%%%%%%%%%%%%%%%%%%%%%%%%%%%%%%%%%%%%%%%%%%
\begin{figure}[h]
\centerline{{\bf a)} 
\psfig{figure=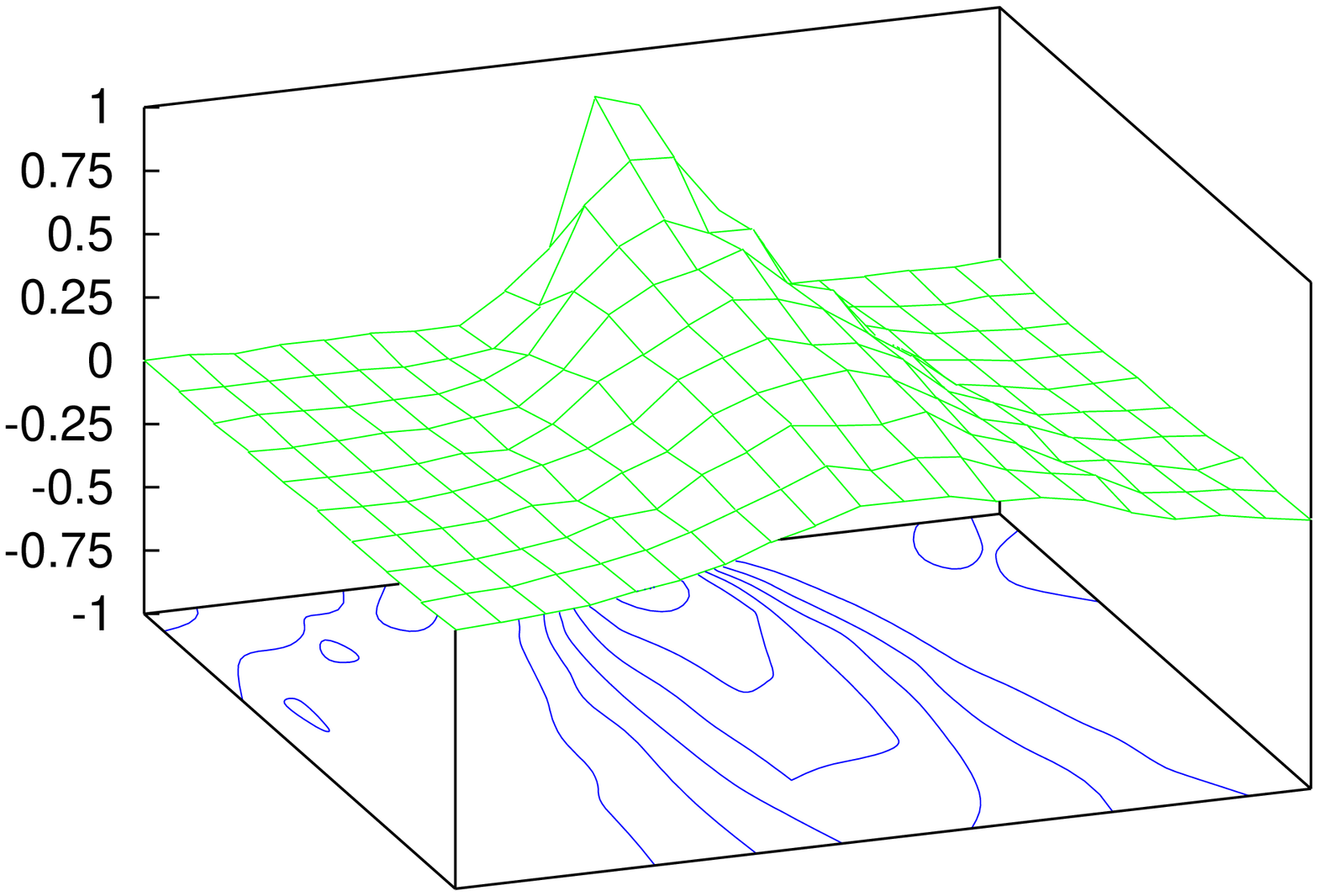,width=8cm,angle=270}
}\centerline{{\bf b)} 
\psfig{figure=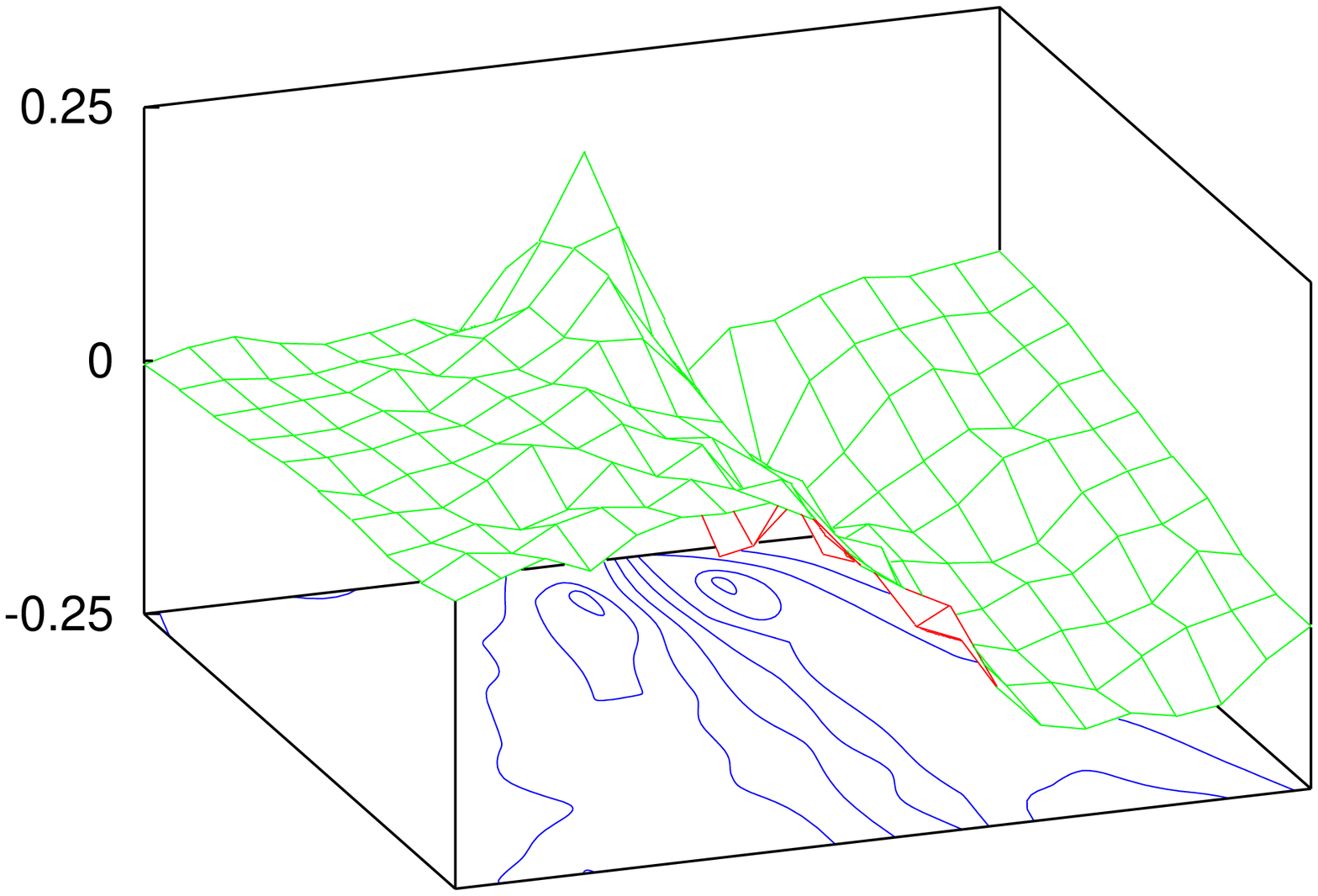,width=8cm,angle=270}
}
\caption{{} Response functions experimentally obtained after averaging over 
  330 packings of 400 disks. The displacement displacement response measures
  the displacement (\textbf{(a)} vertical, \textbf{(b)} horizontal) of a disk
  at $\{x,y\}$ in the bulk, produced by the upwards motion of a disk at
  $\{0,0\}$ in the lowermost layer.  On isostatic systems this is exactly
  equal to the the the vertical excess force on the lowermost central grain at
  $\{0,0\}$ when a (respectively vertical or horizontal) unit force is applied
  at $\{x,y\}$.  }
\label{fig:eresponses}
\end{figure}
%%%%%%%%%%%%%%%%%%%%%%%%%%%%%%%%%%%%%%%%%%%%%%%%%%%%%%%%%%%%%%%%%%%%%%%%%%%%%
\\
Response functions were averaged over 330 repetitions of the experiment. In
order to take averages we subdivide the image into square cells, adding values
of $G$ only to the cell to which the center of the corresponding disk belongs
before the displacement.
\subsection{Experimental results}
\label{sec:eresults}
\begin{figure}[h]
  \centerline{{\bf a)} \psfig{figure=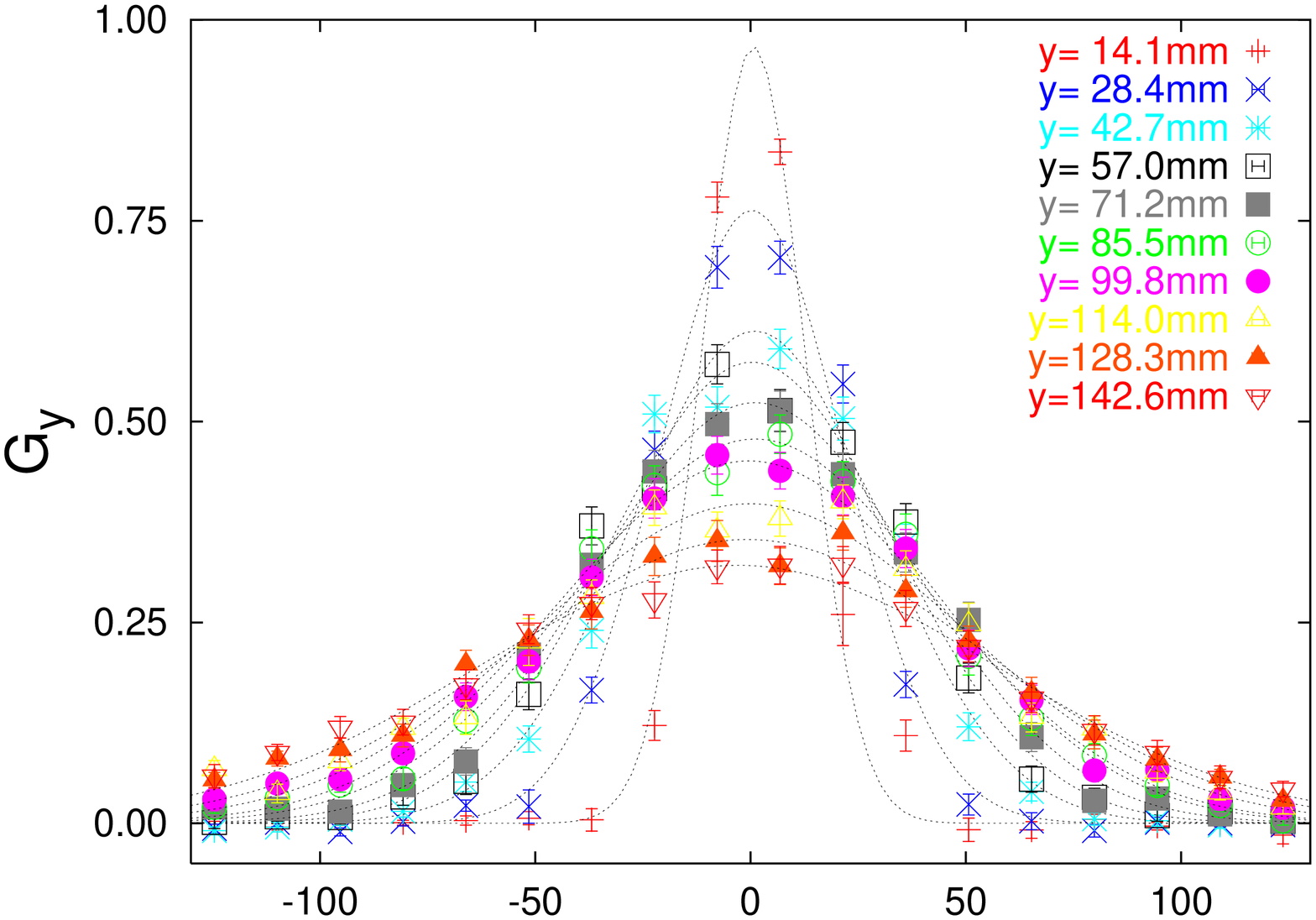,width=8.4cm,angle=270}}
  \centerline{{\bf b)} \psfig{figure=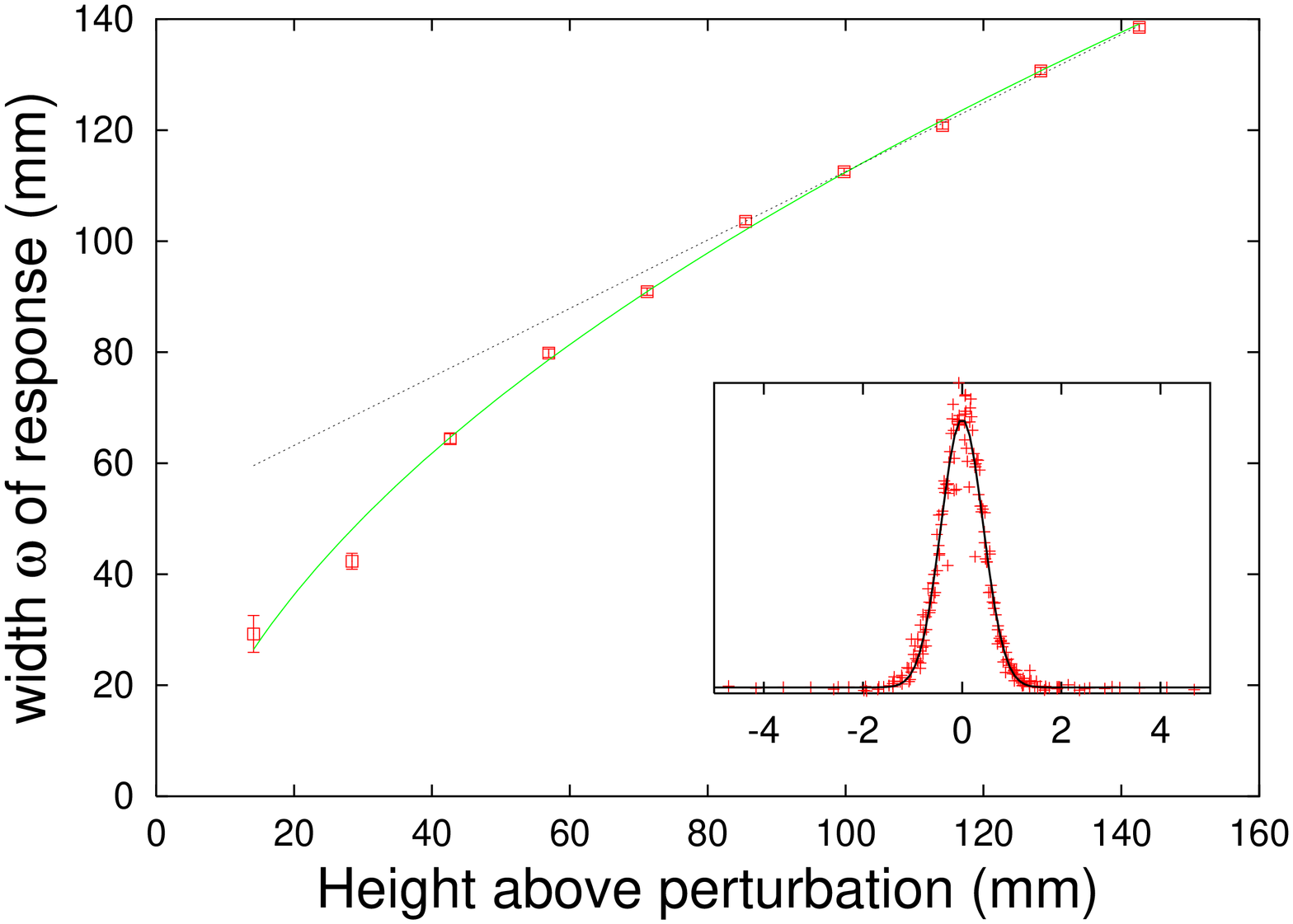,width=8.4cm,angle=270}}
\caption{{} {\bf a)} Experimentally measured vertical response function $G_y$
  at different heights $y$ above the bottom. The dashed lines are Gaussian
  fits.  {\bf b)} Width $\omega$ of $G_y(x,y)$ vs height $y$ above the
  perturbation point (squares). The solid line is a parabolic fit $w(y) = (2
  Dy)^{1/2} + c$ resulting in $D=95 \pm 5$ mm. The dotted straight line has
  slope $0.62$, and fits the last five points. The inset shows the rescaled
  response function $\hat G = \omega G(x/\omega,y)$, for all values of $y$.
  The solid line is a Gaussian fit.}
\label{fig:width}
\end{figure}

\Fig{fig:eresponses}a,b show our main results, respectively vertical and
horizontal response functions. The horizontal response $G_x$ equals minus the
excess compressive force at the bottom produced by a unit force acting in the
positive $x$ direction in the bulk. \Fig{fig:eresponses}b indicates that, on
average, compressive forces on the bottom increase on the right side of the
load's application point, and decrease to the left of it. The $q$-model has no
prediction for this quantity, as it only handles the vertical component of the
force. It would be interesting to compare these results with the predictions
of other competing stress-transmission theories.
\\
As \Fig{fig:width}a shows, the vertical response $G_y$ has the form of a
bell-shaped curve with a single peak, whose width grows with the distance from
the perturbation point. Two models that predict a single-peaked response are
the $q$-model~\cite{LNSF95,CLMM96,D-SRS00}, and classical elasticity. Within
linear elasticity, the width $\omega(y)$ of the response grows linearly with
distance, while in diffusive models like the $q$-model it grows with the
square root of distance.
\\
The width at half-height $\omega(y)$ of the vertical response function $G_y$
can be calculated
% as \hbox{$w(y) = 2 (2\ln 2)^{1/2}(<x^2>_y - <x>_y^2))^{1/2}$}, where $<x^n>_y
% = \int dx x^n G_y(x,y) dx/\int dx G_y(x,y)$.
from the data in \Fig{fig:eresponses}a. Our results are displayed in
\Fig{fig:width}b.  Taking $a,b,c$ as free parameters we fit $\omega(y)
=a*y^b+c$ and obtain $b=0.51 \pm 0.08$. This result, taken at face value,
supports diffuse behavior of stresses.  Assuming diffusive behavior ($b=1/2$)
and fitting $w(y) = (2 Dy)^{1/2} + c$, we obtain $D=95\pm 5$mm. This last
result is not in very good agreement with the diffusive $q$-model theory,
which predicts~\cite{D-SRS00} $D\approx$ grain size/2.  Given that we have
equal numbers of disks of size $16, 17, 18$ and $19$ mm, the average size is
$\approx 17.5$ mm. Thus we find $D \approx 5 \times$ grain size, a factor of
10 of from the theoretical prediction.  With this evidence, we believe that a
parabolic \emph{per se} fit cannot be taken as strong evidence in favor of
diffusive behavior at large scales, given the small number of layers (around
10) that we have in this experiment~\footnote{Similar caveats apply to the
  results reported in Ref.~\cite{D-SRS00}}. Notice that an asymptotically
linear behavior of $\omega(y)$ is also consistent with our data (dotted line
in \ref{fig:width}), if deviations in the first few layers are ignored.  Our
preliminary conclusion is then that parabolic widening holds on very short
scales, however with a possible crossover to linear widening on larger scales.
%%%%%%%%%%%%%%%%%%%%%%%%%%%%%%%%%%%%%%%%%%%%%%%%%%%%%%%%%%%%%%%%%%%%%%%%%%%%%
%%%%%%%%%%%%%%%%%%%%%%%%%%%%%%%%%%%%%%%%%%%%%%%%%%%%%%%%%%%%%%%%%%%%%%%%%%%%%%
\\
In Ref.~\cite{D-SRS00} it is argued that an isostatic system of frictionless
disks should behave according to the predictions of the $q$-model, i.e.\ 
diffusively. This expectation is not verified in
simulations~\cite{MR02,MI01,MI98} in which the polydispersity is small and all
disk centers are located on the sites of a triangular lattice (notice that,
although disk centers are on a regular lattice, these systems have
\emph{strong} contact disorder). For these (on-lattice) isostatic systems, the
average response function of frictionless hard disks shows two peaks that
diverge linearly. This behavior is consistent with theoretical
arguments~\cite{TWS99} suggesting that the assumptions leading to wave-like
propagation~\cite{BCCS95} of stresses are exact on isostatic~\cite{MI98}
packings.
\\
However it has been argued that the response might become single-peaked when
the disorder is large~\cite{GHLF01}.  In order to explore the effect of
disorder in the positions of the disks, and for the sake of comparison with
our own experiments, we performed numerical simulations to measure the
response on systems of frictionless disks with the same distribution of radii
as in the experiments. In these simulations, disks do not occupy the sites of
a regular lattice. This is the subject of next section.
\section{Numerical Experiments}
\label{sec:numer-exper}
The numerical experiments start by pouring disks, one by one, into a
rectangular die, following a steepest descent algorithm~\cite{VKHS01,VIMF03}.
The equilibrium position for each grain is attained when its center of mass
falls between the centers of two already deposited grains. This way of packing
originates a sequentially deposited isostatic structure, however not
necessarily a stable one (positive stresses) as is the case for the on-lattice
simulations of Refs.~\cite{MI98,MI01,MR02}, or the adaptive simulations in
Refs.~\cite{TWS00,HTWR01,HTWR02}. The geometric parameters (size and number of
disks, size of the container, etc.) used in the simulations reproduce the
scale of the real experiment previously discussed. There is no friction in our
simulations.
\\
Once the assembly is ready, the central particle at the bottom of the die is
displaced upwards. Upon perturbing the system we do not allow for
rearrangements, that is, we keep the list of contacts unchanged.  The
isostaticity of the contact network then allows one to calculate the
displacement of all other particles very straightforwardly by upwards
propagation~\cite{MI98,MI01,MR02,VIMF03}.  Because of the conservation of
contacts our results are relevant for the limit of very small perturbations in
frictionless systems. In experiments, rearrangements are very difficult to
avoid when perturbing the system as we do, unless displacements are
exceedingly small.
%%%%%%%%%%%%%%%%%%%%%%%%%%%%%%%%%%%%%%%%%%%%%%%%%%%%%%%%%%%%%%%%%%%%%%%
\begin{figure}[h]
\centerline{{\bf a)} 
\psfig{figure=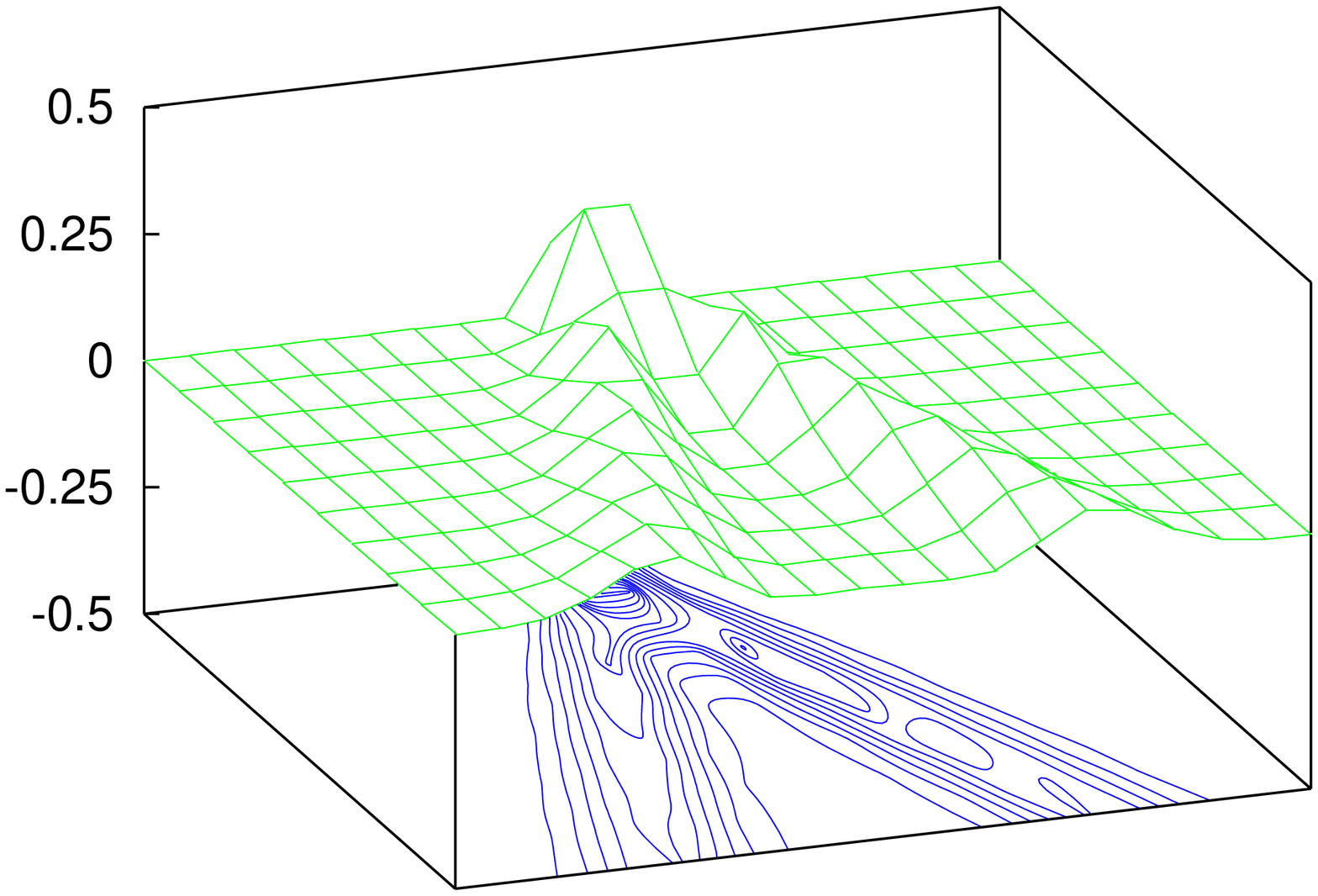,width=8cm,angle=270}
}
\centerline{{\bf b)} 
\psfig{figure=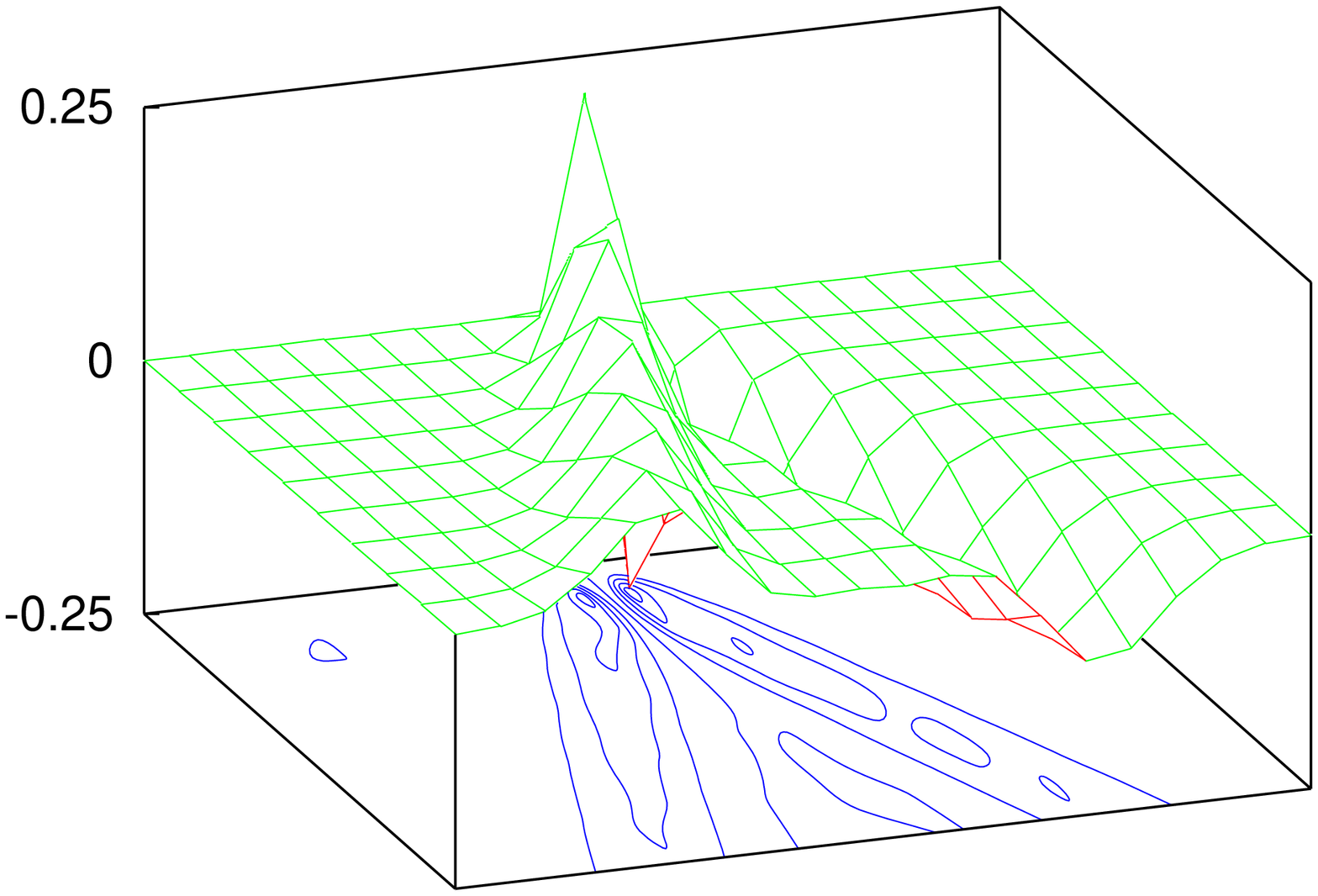,width=8cm,angle=270}
}
\caption{{} Numerical results for the vertical (\textbf{a)}) and horizontal
  (\textbf{b)}) response function, on model frictionless packings with the
  same geometry as in experiments. These responses are calculated assuming the
  limit of infinitesimally small perturbation, in which contacts between disks
  do not change.}
\label{fig:numresponses}
\end{figure}
%%%%%%%%%%%%%%%%%%%%%%%%%%%%%%%%%%%%%%%%%%%%%%%%%%%%%%%%%%%%%%%%%%%%%%%%%%%%
\\
Because of the existence of multiplicative effects~\cite{MR02} in the
simulated experiment, we find the expected large fluctuations in the measured
response functions, thus averages are taken over $10^7$ samples.
\Fig{fig:numresponses} shows the results for the response functions, as
measured in our simulations. It is clear that, also in this case as for
regular packings~\cite{MI98,MI01,MR02} $G_y$ presents a double-peaked shape,
characteristic of a hyperbolic~\cite{BCCS95,CBCM98,HTWR01} behavior. Thus, the
amount of disorder considered in this work does not produce a single-peaked
response in isostatic systems, although the equivalent but frictional
experimental system (\Fig{fig:eresponses}) displays a single peak.  Comparable
results are found with a more elaborate but time-consuming adaptive
algorithm~\cite{TWS00,HTWR01,HTWR02}. 
\section{Discussion}
\label{sec:discussion}
We have measured displacement response functions, both experimentally on
arrays of frictional disks, and numerically for disks without friction.
Because of the virtual work principle, for isostatic systems the displacement
response equals the stress response. Our simulations do have this exact
symmetry, however in experiments the existence of friction makes the system
not isostatic in general~\cite{SEGG02}. Thus the displacement response is not
necessarily equal to the stress green function, however we expect the
differences between them to be small. In this work we have then assumed that
stress responses take similar values to what we find experimentally for
displacement responses. The experimentally measured response
(Figs.~\ref{fig:eresponses} and \ref{fig:width}) has a Gaussian-like shape,
and its width scales approximately as the square root of the depth.  This
appears as consistent with predictions of the
$q$-model~\cite{LNSF95,CLMM96,D-SRS00}.  However, this model predicts a
diffusion coefficient $D$ whose value is of order half the average particle
size. The diffusion coefficient $D$ that we obtain from fitting our
experimental data (\Fig{fig:width}) is ten times larger than this prediction.
We notice that, up to now, diffusive behavior has been clearly seen only on
experimental systems of no more than ten layers.  We must thus remark that our
results are also consistent with a crossover to linear broadening at large
scales. The shape of the response is better approximated by a Gaussian than by
a Lorentzian, as isotropic elasticity would imply.  Recent experiments on
sand~\cite{SRCS01,RCG01} find linear broadening of the response on scales of
the order of 100 layers. Similarly in those experiments the precise shape of
the response is not Lorentzian.  Photoelastic
experiments~\cite{GHLF01,GRCG03x} on somewhat smaller piles also suggest
linear broadening.
\\
Clearly, larger systems must be studied before stronger conclusions can be
drawn from experiments like the one reported in this work. Experiments on more
extended systems are under course at present. However in view of the present
results, as well as those of previous
investigations~\cite{D-SRS00,SRCS01,RCG01,GHLF01,GRCG03x}, a possible scenario
is to have diffusive behavior at short scales, crossing over to some sort of
effective linear elastic behavior (with linear broadening of response) at
larger scales. This would be in line with the expectation that linear
elasticity should be essentially correct at large enough scales.
\\
Our numerical results, on the other hand, show wave-like propagation of
stresses, evidenced by two diverging peaks in the response function. There is
no sign of crossover to a single-peaked response. This is consistent with
previous results~\cite{TWS00,HTWR01,HTWR02}, and confirms that, on
frictionless isostatic systems, disorder does not produce a single-peaked
response. Our numerical results also show that the simplifying assumptions
leading to the $q$-model are not justifiable for frictionless polydisperse
systems.
\\
We remark that our numerical results are valid for the infinitesimal response
function, i.e.\ when contact rearrangements can be
ignored~\cite{RG00,RC02,HTWR02}.  In practice the limit of infinitesimally
small perturbation may be very difficult to attain in experiments, so it would
be desirable to have a means to quantify the effect of rearrangements in
numerical simulations of responses.  Recent experiments study the effect of
rearrangements in packings subject to relatively large
perturbation~\cite{KCLR03x}.
\\
On the other hand, in most experimentally feasible setups, the existence of
friction partially removes the isostaticity properties of granular
packings~\cite{SEGG02}. In the presence of friction the system of equations in
terms of interparticle forces becomes indeterminate~\cite{RBRN96,HEA99} and
accepts a multiplicity of solutions.  Thus friction may be seen as an
additional source of randomness (apart from geometric and contact disorder).
Numerical consideration of friction effects normally makes simulations very
time-consuming~\cite{RBRN96,HEA99,BCCS02}. Clearly it would be interesting to
include the effect of contact rearrangements, and friction, in a realistic but
efficient way in simulations.
\acknowledgements
A.M.V. wishes to thank the Applied Physics Department of CINVESTAV M\'erida,
where this work was done, for hospitality and support.  This work was
partially supported by CONACYT, M\'exico, through research project
\hbox{36256-E}.
\bibliography{}

\begin{thebibliography}{10}
\expandafter\ifx\csname bibnamefont\endcsname\relax
  \def\bibnamefont#1{#1}\fi
\expandafter\ifx\csname bibfnamefont\endcsname\relax
  \def\bibfnamefont#1{#1}\fi
\expandafter\ifx\csname url\endcsname\relax
  \def\url#1{\texttt{#1}}\fi
\expandafter\ifx\csname urlprefix\endcsname\relax\def\urlprefix{URL }\fi
\expandafter\ifx\csname bibinfo\endcsname\relax \def\bibinfo#1#2{#2}\fi
\expandafter\ifx\csname eprint\endcsname\relax \def\eprint#1{#1}\fi

\bibitem{JNBG96}
\bibinfo{author}{\bibfnamefont{H.}~\bibnamefont{Jaeger}},
  \bibinfo{author}{\bibfnamefont{S.}~\bibnamefont{Nagel}}, \bibnamefont{and}
  \bibinfo{author}{\bibfnamefont{R.}~\bibnamefont{Behringer}},
  \bibinfo{journal}{Rev. Mod. Phys.}
  \textbf{\bibinfo{volume}{68}}(\bibinfo{number}{4}), \bibinfo{pages}{1259}
  (\bibinfo{year}{1996}).

\bibitem{WGF97}
\bibinfo{editor}{\bibfnamefont{D.~E.} \bibnamefont{Wolf}} \bibnamefont{and}
  \bibinfo{editor}{\bibfnamefont{P.}~\bibnamefont{Grassberger}}, eds.,
  \emph{\bibinfo{title}{Friction, Arching and Contact Dynamics}}
  (\bibinfo{publisher}{World Scientific}, \bibinfo{address}{Singapore},
  \bibinfo{year}{1997}).

\bibitem{GR98}
\bibinfo{author}{\bibfnamefont{P.}~\bibnamefont{de~Gennes}},
  \bibinfo{journal}{Physica A}
  \textbf{\bibinfo{volume}{261}}(\bibinfo{number}{3-4}), \bibinfo{pages}{267}
  (\bibinfo{year}{1998}).

\bibitem{GG99b}
\bibinfo{author}{\bibfnamefont{P.}~\bibnamefont{de~Gennes}},
  \bibinfo{journal}{Rev. Mod. Phys.} \textbf{\bibinfo{volume}{71}},
  \bibinfo{pages}{S374} (\bibinfo{year}{1999}), ISSN \bibinfo{issn}{0034-6861}.

\bibitem{LNSF95}
\bibinfo{author}{\bibfnamefont{C.}~\bibnamefont{Liu}},
  \bibinfo{author}{\bibfnamefont{S.}~\bibnamefont{Nagel}},
  \bibinfo{author}{\bibfnamefont{D.}~\bibnamefont{Schecter}},
  \bibinfo{author}{\bibfnamefont{S.}~\bibnamefont{Coppersmith}},
  \bibinfo{author}{\bibfnamefont{S.}~\bibnamefont{Majumdar}},
  \bibinfo{author}{\bibfnamefont{O.}~\bibnamefont{Narayan}}, \bibnamefont{and}
  \bibinfo{author}{\bibfnamefont{T.}~\bibnamefont{Witten}},
  \bibinfo{journal}{Science}
  \textbf{\bibinfo{volume}{269}}(\bibinfo{number}{5223}), \bibinfo{pages}{513}
  (\bibinfo{year}{1995}).

\bibitem{CLMM96}
\bibinfo{author}{\bibfnamefont{S.}~\bibnamefont{Coppersmith}},
  \bibinfo{author}{\bibfnamefont{C.}~\bibnamefont{Liu}},
  \bibinfo{author}{\bibfnamefont{S.}~\bibnamefont{Majumdar}},
  \bibinfo{author}{\bibfnamefont{O.}~\bibnamefont{Narayan}}, \bibnamefont{and}
  \bibinfo{author}{\bibfnamefont{T.}~\bibnamefont{Witten}},
  \bibinfo{journal}{Phys. Rev. E}
  \textbf{\bibinfo{volume}{53}}(\bibinfo{number}{5}), \bibinfo{pages}{4673}
  (\bibinfo{year}{1996}).

\bibitem{SSSF99}
\bibinfo{author}{\bibfnamefont{M.}~\bibnamefont{Sexton}},
  \bibinfo{author}{\bibfnamefont{J.}~\bibnamefont{Socolar}}, \bibnamefont{and}
  \bibinfo{author}{\bibfnamefont{D.}~\bibnamefont{Schaeffer}},
  \bibinfo{journal}{Phys. Rev. E}
  \textbf{\bibinfo{volume}{60}}(\bibinfo{number}{2}), \bibinfo{pages}{1999}
  (\bibinfo{year}{1999}).

\bibitem{CBS97}
\bibinfo{author}{\bibfnamefont{P.}~\bibnamefont{Claudin}} \bibnamefont{and}
  \bibinfo{author}{\bibfnamefont{J.}~\bibnamefont{Bouchaud}},
  \bibinfo{journal}{Phys. Rev. Lett.}
  \textbf{\bibinfo{volume}{78}}(\bibinfo{number}{2}), \bibinfo{pages}{231}
  (\bibinfo{year}{1997}).

\bibitem{BCCS95}
\bibinfo{author}{\bibfnamefont{J.}~\bibnamefont{Bouchaud}},
  \bibinfo{author}{\bibfnamefont{M.}~\bibnamefont{Cates}}, \bibnamefont{and}
  \bibinfo{author}{\bibfnamefont{P.}~\bibnamefont{Claudin}},
  \bibinfo{journal}{J. Phys. I}
  \textbf{\bibinfo{volume}{5}}(\bibinfo{number}{6}), \bibinfo{pages}{639}
  (\bibinfo{year}{1995}).

\bibitem{WCCA96}
\bibinfo{author}{\bibfnamefont{J.}~\bibnamefont{Wittmer}},
  \bibinfo{author}{\bibfnamefont{P.}~\bibnamefont{Claudin}},
  \bibinfo{author}{\bibfnamefont{M.}~\bibnamefont{Cates}}, \bibnamefont{and}
  \bibinfo{author}{\bibfnamefont{J.}~\bibnamefont{Bouchaud}},
  \bibinfo{journal}{Nature}
  \textbf{\bibinfo{volume}{382}}(\bibinfo{number}{6589}), \bibinfo{pages}{336}
  (\bibinfo{year}{1996}).

\bibitem{WCCS97}
\bibinfo{author}{\bibfnamefont{J.}~\bibnamefont{Wittmer}},
  \bibinfo{author}{\bibfnamefont{M.}~\bibnamefont{Cates}}, \bibnamefont{and}
  \bibinfo{author}{\bibfnamefont{P.}~\bibnamefont{Claudin}},
  \bibinfo{journal}{J. Phys. I}
  \textbf{\bibinfo{volume}{7}}(\bibinfo{number}{1}), \bibinfo{pages}{39}
  (\bibinfo{year}{1997}).

\bibitem{VCBS00}
\bibinfo{author}{\bibfnamefont{L.}~\bibnamefont{Vanel}},
  \bibinfo{author}{\bibfnamefont{P.}~\bibnamefont{Claudin}},
  \bibinfo{author}{\bibfnamefont{J.}~\bibnamefont{Bouchaud}},
  \bibinfo{author}{\bibfnamefont{M.}~\bibnamefont{Cates}},
  \bibinfo{author}{\bibfnamefont{E.}~\bibnamefont{Clement}}, \bibnamefont{and}
  \bibinfo{author}{\bibfnamefont{J.}~\bibnamefont{Wittmer}},
  \bibinfo{journal}{Phys. Rev. Lett.}
  \textbf{\bibinfo{volume}{84}}(\bibinfo{number}{7}), \bibinfo{pages}{1439}
  (\bibinfo{year}{2000}).

\bibitem{KSPN98}
\bibinfo{author}{\bibfnamefont{V.}~\bibnamefont{Kenkre}},
  \bibinfo{author}{\bibfnamefont{J.}~\bibnamefont{Scott}},
  \bibinfo{author}{\bibfnamefont{E.}~\bibnamefont{Pease}}, \bibnamefont{and}
  \bibinfo{author}{\bibfnamefont{A.}~\bibnamefont{Hurd}},
  \bibinfo{journal}{Phys. Rev. E}
  \textbf{\bibinfo{volume}{57}}(\bibinfo{number}{5}), \bibinfo{pages}{5841}
  (\bibinfo{year}{1998}).

\bibitem{KS01}
\bibinfo{author}{\bibnamefont{{Kenkre, VM}}}, \bibinfo{journal}{Granul. Matter}
  \textbf{\bibinfo{volume}{3}}(\bibinfo{number}{1-2}), \bibinfo{pages}{23}
  (\bibinfo{year}{2001}).

\bibitem{BCLF01}
\bibinfo{author}{\bibfnamefont{J.}~\bibnamefont{Bouchaud}},
  \bibinfo{author}{\bibfnamefont{P.}~\bibnamefont{Claudin}},
  \bibinfo{author}{\bibfnamefont{D.}~\bibnamefont{Levine}}, \bibnamefont{and}
  \bibinfo{author}{\bibfnamefont{M.}~\bibnamefont{Otto}},
  \bibinfo{journal}{Eur. Phys. J. E}
  \textbf{\bibinfo{volume}{4}}(\bibinfo{number}{4}), \bibinfo{pages}{451}
  (\bibinfo{year}{2001}).

\bibitem{D-SRS00}
\bibinfo{author}{\bibfnamefont{M.}~\bibnamefont{Da~Silva}} \bibnamefont{and}
  \bibinfo{author}{\bibfnamefont{J.}~\bibnamefont{Rajchenbach}},
  \bibinfo{journal}{Nature}
  \textbf{\bibinfo{volume}{406}}(\bibinfo{number}{6797}), \bibinfo{pages}{708}
  (\bibinfo{year}{2000}).

\bibitem{GRCG03x}
\bibinfo{author}{\bibfnamefont{J.}~\bibnamefont{Geng}},
  \bibinfo{author}{\bibfnamefont{G.}~\bibnamefont{Reydellet}},
  \bibinfo{author}{\bibfnamefont{E.}~\bibnamefont{{Cl\'ement}}},
  \bibnamefont{and} \bibinfo{author}{\bibfnamefont{R.~P.}
  \bibnamefont{Behringer}}, \emph{\bibinfo{title}{Green's Function measurements
  of force transmission in 2D granular materials}} (\bibinfo{year}{2003}),
  \eprint{arXiv:condmat/0211031v3}.

\bibitem{SKHN98}
\bibinfo{author}{\bibfnamefont{J.}~\bibnamefont{Scott}},
  \bibinfo{author}{\bibfnamefont{V.}~\bibnamefont{Kenkre}}, \bibnamefont{and}
  \bibinfo{author}{\bibfnamefont{A.}~\bibnamefont{Hurd}},
  \bibinfo{journal}{Phys. Rev. E}
  \textbf{\bibinfo{volume}{57}}(\bibinfo{number}{5}), \bibinfo{pages}{5850}
  (\bibinfo{year}{1998}).

\bibitem{RCG01}
\bibinfo{author}{\bibfnamefont{G.}~\bibnamefont{Reydellet}} \bibnamefont{and}
  \bibinfo{author}{\bibfnamefont{E.}~\bibnamefont{Clement}},
  \bibinfo{journal}{Phys. Rev. Lett.}
  \textbf{\bibinfo{volume}{86}}(\bibinfo{number}{15}), \bibinfo{pages}{3308}
  (\bibinfo{year}{2001}).

\bibitem{SRCS01}
\bibinfo{author}{\bibfnamefont{D.}~\bibnamefont{Serero}},
  \bibinfo{author}{\bibfnamefont{G.}~\bibnamefont{Reydellet}},
  \bibinfo{author}{\bibfnamefont{P.}~\bibnamefont{Claudin}},
  \bibinfo{author}{\bibfnamefont{E.}~\bibnamefont{Clement}}, \bibnamefont{and}
  \bibinfo{author}{\bibfnamefont{D.}~\bibnamefont{Levine}},
  \bibinfo{journal}{Eur. Phys. J. E}
  \textbf{\bibinfo{volume}{6}}(\bibinfo{number}{2}), \bibinfo{pages}{169}
  (\bibinfo{year}{2001}).

\bibitem{OBCA03}
\bibinfo{author}{\bibfnamefont{M.}~\bibnamefont{Otto}},
  \bibinfo{author}{\bibfnamefont{J.}~\bibnamefont{Bouchaud}},
  \bibinfo{author}{\bibfnamefont{P.}~\bibnamefont{Claudin}}, \bibnamefont{and}
  \bibinfo{author}{\bibfnamefont{J.}~\bibnamefont{Socolar}},
  \bibinfo{journal}{Phys. Rev. E}
  \textbf{\bibinfo{volume}{67}}(\bibinfo{number}{3}), \bibinfo{pages}{031302}
  (\bibinfo{year}{2003}).

\bibitem{MI98}
\bibinfo{author}{\bibnamefont{{Moukarzel, CF}}}, \bibinfo{journal}{Phys. Rev.
  Lett.} \textbf{\bibinfo{volume}{81}}(\bibinfo{number}{8}),
  \bibinfo{pages}{1634} (\bibinfo{year}{1998}).

\bibitem{RG00}
\bibinfo{author}{\bibnamefont{{Roux, JN}}}, \bibinfo{journal}{Phys. Rev. E}
  \textbf{\bibinfo{volume}{61}}(\bibinfo{number}{6}), \bibinfo{pages}{6802}
  (\bibinfo{year}{2000}).

\bibitem{MI01}
\bibinfo{author}{\bibnamefont{{Moukarzel, CF}}}, \bibinfo{journal}{Granul.
  Matter} \textbf{\bibinfo{volume}{3}}(\bibinfo{number}{1-2}),
  \bibinfo{pages}{41} (\bibinfo{year}{2001}).

\bibitem{MR02}
\bibinfo{author}{\bibnamefont{{Moukarzel, CF}}}, \bibinfo{journal}{J.
  Phys.-Condes. Matter} \textbf{\bibinfo{volume}{14}}(\bibinfo{number}{9}),
  \bibinfo{pages}{2379} (\bibinfo{year}{2002}).

\bibitem{TWS99}
\bibinfo{author}{\bibfnamefont{A.}~\bibnamefont{Tkachenko}} \bibnamefont{and}
  \bibinfo{author}{\bibfnamefont{T.}~\bibnamefont{Witten}},
  \bibinfo{journal}{Phys. Rev. E}
  \textbf{\bibinfo{volume}{60}}(\bibinfo{number}{1}), \bibinfo{pages}{687}
  (\bibinfo{year}{1999}).

\bibitem{TWS00}
\bibinfo{author}{\bibfnamefont{A.}~\bibnamefont{Tkachenko}} \bibnamefont{and}
  \bibinfo{author}{\bibfnamefont{T.}~\bibnamefont{Witten}},
  \bibinfo{journal}{Phys. Rev. E}
  \textbf{\bibinfo{volume}{62}}(\bibinfo{number}{2}), \bibinfo{pages}{2510}
  (\bibinfo{year}{2000}).

\bibitem{HTWR01}
\bibinfo{author}{\bibfnamefont{D.}~\bibnamefont{Head}},
  \bibinfo{author}{\bibfnamefont{A.}~\bibnamefont{Tkachenko}},
  \bibnamefont{and} \bibinfo{author}{\bibfnamefont{T.}~\bibnamefont{Witten}},
  \bibinfo{journal}{Eur. Phys. J. E}
  \textbf{\bibinfo{volume}{6}}(\bibinfo{number}{1}), \bibinfo{pages}{99}
  (\bibinfo{year}{2001}).

\bibitem{GHLF01}
\bibinfo{author}{\bibfnamefont{J.}~\bibnamefont{Geng}},
  \bibinfo{author}{\bibfnamefont{D.}~\bibnamefont{Howell}},
  \bibinfo{author}{\bibfnamefont{E.}~\bibnamefont{Longhi}},
  \bibinfo{author}{\bibfnamefont{R.}~\bibnamefont{Behringer}},
  \bibinfo{author}{\bibfnamefont{G.}~\bibnamefont{Reydellet}},
  \bibinfo{author}{\bibfnamefont{L.}~\bibnamefont{Vanel}},
  \bibinfo{author}{\bibfnamefont{E.}~\bibnamefont{Clement}}, \bibnamefont{and}
  \bibinfo{author}{\bibfnamefont{S.}~\bibnamefont{Luding}},
  \bibinfo{journal}{Phys. Rev. Lett.}
  \textbf{\bibinfo{volume}{8703}}(\bibinfo{number}{3}), \bibinfo{pages}{5506}
  (\bibinfo{year}{2001}).

\bibitem{RC02}
\bibinfo{author}{\bibnamefont{{Roux, JN}}}, \bibinfo{journal}{Eur. Phys. J. E}
  \textbf{\bibinfo{volume}{7}}(\bibinfo{number}{3}), \bibinfo{pages}{297}
  (\bibinfo{year}{2002}).

\bibitem{HTWR02}
\bibinfo{author}{\bibfnamefont{D.}~\bibnamefont{Head}},
  \bibinfo{author}{\bibfnamefont{A.}~\bibnamefont{Tkachenko}},
  \bibnamefont{and} \bibinfo{author}{\bibfnamefont{T.}~\bibnamefont{Witten}},
  \bibinfo{journal}{Eur. Phys. J. E}
  \textbf{\bibinfo{volume}{7}}(\bibinfo{number}{3}), \bibinfo{pages}{299}
  (\bibinfo{year}{2002}).

\bibitem{RBRN96}
\bibinfo{author}{\bibfnamefont{F.}~\bibnamefont{Radjai}},
  \bibinfo{author}{\bibfnamefont{L.}~\bibnamefont{Brendel}}, \bibnamefont{and}
  \bibinfo{author}{\bibfnamefont{S.}~\bibnamefont{Roux}},
  \bibinfo{journal}{Phys. Rev. E}
  \textbf{\bibinfo{volume}{54}}(\bibinfo{number}{1}), \bibinfo{pages}{861}
  (\bibinfo{year}{1996}).

\bibitem{HEA99}
\bibinfo{author}{\bibfnamefont{T.}~\bibnamefont{Halsey}} \bibnamefont{and}
  \bibinfo{author}{\bibfnamefont{D.}~\bibnamefont{Ertas}},
  \bibinfo{journal}{Phys. Rev. Lett.}
  \textbf{\bibinfo{volume}{83}}(\bibinfo{number}{24}), \bibinfo{pages}{5007}
  (\bibinfo{year}{1999}).

\bibitem{SEGG02}
\bibinfo{author}{\bibfnamefont{L.}~\bibnamefont{Silbert}},
  \bibinfo{author}{\bibfnamefont{D.}~\bibnamefont{Ertas}},
  \bibinfo{author}{\bibfnamefont{G.}~\bibnamefont{Grest}},
  \bibinfo{author}{\bibfnamefont{T.}~\bibnamefont{Halsey}}, \bibnamefont{and}
  \bibinfo{author}{\bibfnamefont{D.}~\bibnamefont{Levine}},
  \bibinfo{journal}{Phys. Rev. E}
  \textbf{\bibinfo{volume}{65}}(\bibinfo{number}{3}), \bibinfo{pages}{031304}
  (\bibinfo{year}{2002}).

\bibitem{MJSP00}
\bibinfo{author}{\bibfnamefont{H.}~\bibnamefont{Makse}},
  \bibinfo{author}{\bibfnamefont{D.}~\bibnamefont{Johnson}}, \bibnamefont{and}
  \bibinfo{author}{\bibfnamefont{L.}~\bibnamefont{Schwartz}},
  \bibinfo{journal}{Phys. Rev. Lett.}
  \textbf{\bibinfo{volume}{84}}(\bibinfo{number}{18}), \bibinfo{pages}{4160}
  (\bibinfo{year}{2000}).

\bibitem{MJNF98}
\bibinfo{author}{\bibfnamefont{D.}~\bibnamefont{Mueth}},
  \bibinfo{author}{\bibfnamefont{H.}~\bibnamefont{Jaeger}}, \bibnamefont{and}
  \bibinfo{author}{\bibfnamefont{S.}~\bibnamefont{Nagel}},
  \bibinfo{journal}{Phys. Rev. E}
  \textbf{\bibinfo{volume}{57}}(\bibinfo{number}{3}), \bibinfo{pages}{3164}
  (\bibinfo{year}{1998}).

\bibitem{BMMF01}
\bibinfo{author}{\bibfnamefont{D.}~\bibnamefont{Blair}},
  \bibinfo{author}{\bibfnamefont{N.}~\bibnamefont{Mueggenburg}},
  \bibinfo{author}{\bibfnamefont{A.}~\bibnamefont{Marshall}},
  \bibinfo{author}{\bibfnamefont{H.}~\bibnamefont{Jaeger}}, \bibnamefont{and}
  \bibinfo{author}{\bibfnamefont{S.}~\bibnamefont{Nagel}},
  \bibinfo{journal}{Phys. Rev. E}
  \textbf{\bibinfo{volume}{6304}}(\bibinfo{number}{4}), \bibinfo{pages}{1304}
  (\bibinfo{year}{2001}).

\bibitem{MJNS02}
\bibinfo{author}{\bibfnamefont{N.}~\bibnamefont{Mueggenburg}},
  \bibinfo{author}{\bibfnamefont{H.}~\bibnamefont{Jaeger}}, \bibnamefont{and}
  \bibinfo{author}{\bibfnamefont{S.}~\bibnamefont{Nagel}},
  \bibinfo{journal}{Phys. Rev. E}
  \textbf{\bibinfo{volume}{66}}(\bibinfo{number}{3}), \bibinfo{pages}{031304}
  (\bibinfo{year}{2002}).

\bibitem{LMFF99}
\bibinfo{author}{\bibfnamefont{G.}~\bibnamefont{Lovoll}},
  \bibinfo{author}{\bibfnamefont{K.}~\bibnamefont{Maloy}}, \bibnamefont{and}
  \bibinfo{author}{\bibfnamefont{E.}~\bibnamefont{Flekkoy}},
  \bibinfo{journal}{Phys. Rev. E}
  \textbf{\bibinfo{volume}{60}}(\bibinfo{number}{5}), \bibinfo{pages}{5872}
  (\bibinfo{year}{1999}).

\bibitem{VKHS01}
\bibinfo{author}{\bibfnamefont{A.}~\bibnamefont{Vidales}},
  \bibinfo{author}{\bibfnamefont{V.}~\bibnamefont{Kenkre}}, \bibnamefont{and}
  \bibinfo{author}{\bibfnamefont{A.}~\bibnamefont{Hurd}},
  \bibinfo{journal}{Granul. Matter}
  \textbf{\bibinfo{volume}{3}}(\bibinfo{number}{1-2}), \bibinfo{pages}{141}
  (\bibinfo{year}{2001}).

\bibitem{VIMF03}
\bibinfo{author}{\bibfnamefont{A.}~\bibnamefont{Vidales}},
  \bibinfo{author}{\bibfnamefont{I.}~\bibnamefont{Ippolito}}, \bibnamefont{and}
  \bibinfo{author}{\bibfnamefont{C.}~\bibnamefont{Moukarzel}},
  \bibinfo{journal}{Physica A}
  \textbf{\bibinfo{volume}{325}}(\bibinfo{number}{3-4}), \bibinfo{pages}{297}
  (\bibinfo{year}{2003}).

\bibitem{CBCM98}
\bibinfo{author}{\bibfnamefont{P.}~\bibnamefont{Claudin}},
  \bibinfo{author}{\bibfnamefont{J.}~\bibnamefont{Bouchaud}},
  \bibinfo{author}{\bibfnamefont{M.}~\bibnamefont{Cates}}, \bibnamefont{and}
  \bibinfo{author}{\bibfnamefont{J.}~\bibnamefont{Wittmer}},
  \bibinfo{journal}{Phys. Rev. E}
  \textbf{\bibinfo{volume}{57}}(\bibinfo{number}{4}), \bibinfo{pages}{4441}
  (\bibinfo{year}{1998}).

\bibitem{KCLR03x}
\bibinfo{author}{\bibfnamefont{E.}~\bibnamefont{Kolb}},
  \bibinfo{author}{\bibfnamefont{J.}~\bibnamefont{Cviklinski}},
  \bibinfo{author}{\bibfnamefont{J.}~\bibnamefont{Lanuza}},
  \bibinfo{author}{\bibfnamefont{P.}~\bibnamefont{Claudin}}, \bibnamefont{and}
  \bibinfo{author}{\bibfnamefont{E.}~\bibnamefont{Clement}},
  \emph{\bibinfo{title}{Reorganization of a dense granular assembly: the
  `unjamming response function'}} (\bibinfo{year}{2003}),
  \eprint{arXiv:cond-mat/0308054}.

\bibitem{BCCS02}
\bibinfo{author}{\bibfnamefont{L.}~\bibnamefont{Breton}},
  \bibinfo{author}{\bibfnamefont{P.}~\bibnamefont{Claudin}},
  \bibinfo{author}{\bibfnamefont{E.}~\bibnamefont{Clement}}, \bibnamefont{and}
  \bibinfo{author}{\bibfnamefont{J.}~\bibnamefont{Zucker}},
  \bibinfo{journal}{Europhys. Lett.}
  \textbf{\bibinfo{volume}{60}}(\bibinfo{number}{6}), \bibinfo{pages}{813}
  (\bibinfo{year}{2002}).

\end{thebibliography}
\end{document}